\documentclass[format=manuscript, review=false, anonymous=false, screen, dvipsnames]{lib-acm/acmart}

\setcopyright{acmlicensed}
\copyrightyear{2018}
\acmYear{2018}
\acmDOI{XXXXXXX.XXXXXXX}

\acmConference[]{}{}{}
\acmYear{}
\copyrightyear{}
\acmPrice{}
\acmDOI{}
\acmISBN{}
\setcopyright{none}

\usepackage{booktabs} %
\usepackage{tabularx}
\usepackage{multirow}
\usepackage{siunitx} 
\usepackage{threeparttable} 
\usepackage{adjustbox}
\usepackage{makecell}
\usepackage{longtable}
\usepackage{xltabular}

\usepackage{exii-macros}

\usepackage{booktabs}

\showrevisions{REVISIONGREEN}

\makeatletter
\g@addto@macro\normalsize{%
  \setlength\abovedisplayshortskip{-9pt}
  \setlength\belowdisplayshortskip{3pt}
}
\makeatother

\begin{document}

\tolerance=400 

\title[Probing How Users Interact with Turn-Level Design Frictions for AI Chatbots]{Probing How Users Interact with Turn-Level Design Frictions for AI Chatbots}

\author{Helen Weixu Chen}
\orcid{0009-0008-1384-6781}
\affiliation{%
  \institution{School of Computer Science\\
  University of Waterloo}
  \country{Waterloo, ON, Canada}
}
\email{w352chen@uwaterloo.ca}

\author{Katy Ilonka Gero}
\orcid{0000-0001-5982-9321}
\affiliation{%
  \institution{School of Computer Science\\
  University of Sydney}
  \country{Sydney, NSW, Australia}
}
\email{katy.gero@sydney.edu.au}

\renewcommand{\shortauthors}{Chen and Gero}

\begin{abstract}
AI chatbots can help people write faster, but they can also encourage overreliance by making it easy to turn minimal input into usable text. We study \textit{turn-level design friction}: intentional constraints added to each chatbot exchange that slow, limit, or redirect how users request, access, or use model responses. We designed six friction probes, organized around three mechanisms: eliciting user contribution, restricting access to generated content, and reshaping system output. In a within-subject study with 24 participants, all six probes increased workload, task duration, and perceived ownership relative to a conventional AI chatbot, while their effects on recall and recognition were more selective. We further found that participants adapted to friction in different ways, and that the same constraint could support or obstruct involvement depending on users' goals and workflows.

\end{abstract}

\ccsdesc[500]{Human-centered computing~Interaction tech}

\keywords{LLM, design friction, ownership, writing, responsible AI, memory}

\begin{teaserfigure}
\centering
  \includegraphics[width=\textwidth]{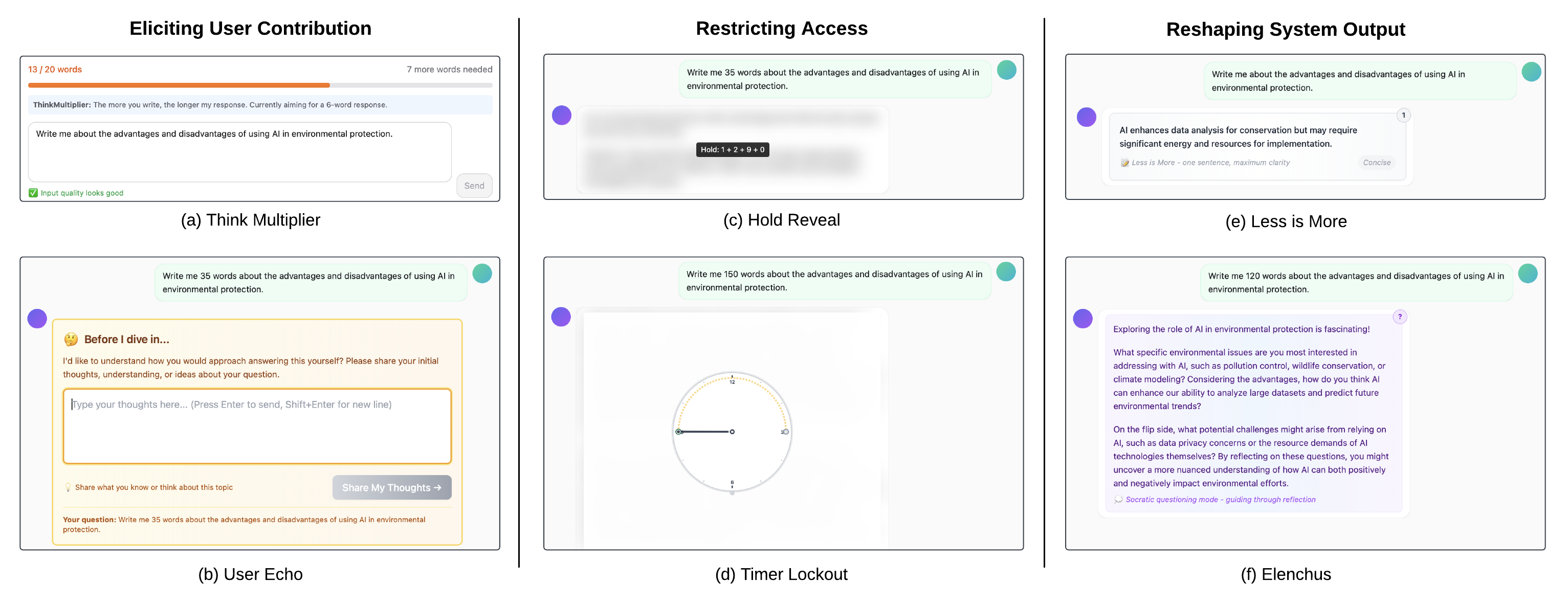}
  \caption{Six friction interfaces, organized into three design mechanisms, explore how AI chatbot interactions can sustain user involvement: eliciting user contribution, restricting access to or reuse of generated content, and reshaping system output: (a) Think Multiplier: Prompt must be $\geq 20$ words; response is $\leq$ half the prompt; (b) User Echo: After submitting a prompt, users draft a tentative response before the model replies; (c) Hold Reveal: Users must hold 1, 2, 9, and 0 simultaneously to reveal the model response; (d) Timer Lockout: Users slide a timer control to reveal the response for the next 15 seconds; (e) Less is More: The model output is limited to one sentence per turn; (f) Elenchus: The model responds only with Socratic-style questions.} 
  \label{fig:teaser}
\end{teaserfigure}

\maketitle

\section{Introduction}
AI chatbots have become a widely used interface for interacting with large language models (LLMs). Their popularity stems from an interaction model that minimizes user effort: users can provide a brief prompt and immediately receive a fluent, usable response. This workflow makes AI assistance valuable across a wide range of activities, including asking questions, brainstorming ideas, learning new concepts, programming, and writing \cite{NBERw34255}. However, the same ease of interaction can also position users as passive consumers of AI-generated content \cite{LIANG2025101366}. Rather than actively developing, evaluating, or refining ideas, users can often move directly from a minimal prompt to a response that is ready to adopt. As AI chatbots become increasingly capable, an important design challenge is therefore not only how to improve model outputs, but also how to preserve meaningful user involvement without losing the practical value of AI assistance.

One promising approach is design friction \cite{frischmann2023friction} --- the intentional introduction of interaction constraints that slow, interrupt, or redirect an otherwise effortless workflow. Friction has been extensively studied in scoial media and digital wellbeing, where it has been used to interrupt habitual use and encourage more deliberate engagement \cite{lukoff2023switchtube, ruiz2024design}. In AI chatbot interactions, friction might require users to contribute more of their own thinking, make model output less immediately reusable, or perform additional actions before accessing it. Prior work has demonstrated the potential value of such interventions. In writing, \citet{joshi2025writing} found that longer prompts were associated with greater psychological ownership and proposed a hold-to-submit delay to encourage users to write longer prompts. In programming education, \citet{kazemitabaar2025exploring} explored cognitive engagement techniques that require learners to explain, trace, or reason through AI-generated code before using it. However, these studies primarily examine individual techniques within particular tasks. We still lack a comparative account of the interaction mechanisms through which friction operates, the forms of effort that different mechanisms require, the costs they introduce, and how users adapt when familiar chatbot workflows are deliberately disrupted.

In this paper, we investigate \textit{turn-level design friction}: interaction constraints embedded within individual chatbot exchanges that change how users request, access, interpret, or use AI-generated responses. Drawing on prior HCI work, we identify three interaction mechanisms for introducing friction: eliciting user contribution, restricting access, and reshaping system output. Guided by these mechanisms, we developed six design probes, with two distinct probes representing each mechanism. We evaluated the probes through a within-subject study with 24 participants, using short writing tasks as a context in which participants had to request, interpret, and incorporate AI-generated material into a concrete response. Our mixed-methods evaluation combined measures of workload, task duration, psychological ownership, recall, and recognition with interaction logs and semi-structured interviews.

Our findings show that introducing additional effort was not sufficient to make an interaction productive. All six friction probes increased workload, task duration, and psychological ownership relative to a conventional AI chatbot, but their effects on recall, recognition, and workflow varied. Across these outcomes, the consequences of friction depended on where the added effort was directed and how the constraint fit users’ goals and workflows. Friction could support involvement when it kept users engaged in formulating, interpreting, or assembling the response, but could instead become obstructive when effort shifted toward satisfying the constraint or managing the interface. Participants also adapted their workflows in response to the constraints, changing how they distributed work between themselves and the model.

We contribute a mechanism-based framing of turn-level design friction, instantiated through six interface probes, and an empirical account of the benefits, costs, and workflow consequences of applying friction to AI chatbot interactions. We provide preliminary evidence for how turn-level frictions can be effective at deepening user engagement, and propose future work to investigate how different mechanisms play out in longer, more naturalistic settings.

\section{Background and Related Work}
\subsection{Impacts of LLM Assistance on User Involvement}
LLMs can substantially reduce the effort required to complete complex tasks, but this convenience can also change how much cognitive work users perform themselves. Prior work has linked over-reliance on AI, such as accepting model recommendations without sufficient scrutiny, to weaker critical thinking, analytical reasoning, and decision-making~\cite{zhai_effects_2024}. Cognitive offloading offers one explanation for these effects: by shifting cognitive processes to external tools, users can reduce the mental work they perform themselves \cite{risko2016cognitive}. Consistent with this account, \citet{soc15010006} found that heavier AI use was associated with lower critical-thinking performance, with cognitive offloading mediating this relationship. 

Beyond reducing the amount of cognitive work users perform, LLM assistance can also shape how they reason about a task. Prior work on AI-assisted decision making has documented automation bias, in which people follow AI recommendations, including incorrect ones, rather than independently evaluating them \cite{buccinca2021trust}. Similar effects appear with generative AI. \citet{choi-etal-2024-llm} found that LLM suggestions increased expert annotation speed by 133.5\%, but also shifted annotators’ decisions toward the model and left several low-prevalence but important concepts overlooked. Even assistance used primarily for navigating information can influence subsequent deliberation: \citet{10.1145/3772318.3791796} found that early LLM access could anchor which information and positions users subsequently considered, with its effects on critical-thinking performance varying with the time available for the task. 

LLM assistance can also change users' involvement in shaping and internalizing the resulting work. In AI-assisted writing, \citet{kreminski2024dearth} argues that when users can produce substantial text while making comparatively few creative decisions, for instance using a short AI prompt to produce a long story, the resulting work may not reflect the author's expressive intent. In knowledge work, \citet{kobiella2024if} similarly found that users' sense of ownership and accomplishment was closely tied to their perceived contribution and control: participants described post-processing ChatGPT output as helping them retain ownership, whereas a more dominant role for ChatGPT could distance them from the resulting work. Neurophysiological evidence points to related differences in how AI-assisted writing is internalized. \citet{kosmyna2025your} observed weaker neural connectivity during LLM-assisted essay writing than during unassisted writing and poorer immediate recall of participants' own essays.

These effects, however, are not inevitable consequences of LLM use. Recent work suggests that how assistance is structured can determine whether it displaces or supports users’ own reasoning. \citet{10.1145/3772318.3791796}, for example, found that beginning a task independently before receiving LLM assistance could support stronger critical-thinking performance when sufficient time was available. Other work shows that structured chatbot interactions can support reflection and problem solving when users are prompted to analyze, explain, and make decisions rather than simply receive answers~\cite{rienovita9utilization}. Overall, prior work suggests that the central design challenge is not simply whether to provide AI assistance, but how to structure that assistance so users remain meaningfully involved in the cognitive work underlying the task.

\subsection{Design Friction}
Design friction challenges the longstanding assumption that interaction should always minimize effort, delay, and interruption \cite{frischmann2023friction, 10.1145/2851581.2892410}. Prior work has argued that small obstacles or ``microboundaries'' can disrupt automatic behaviour and create moments for reflection, shifting interaction from habitual toward more deliberate action \cite{10.1145/2851581.2892410}. This logic has been explored widely in digital wellbeing and social media, through interventions such as goal reminders \cite{lyngs2020just}, time limits and lockouts \cite{10.1145/2851581.2892410}, delayed access \cite{haliburton2024longitudinal}, feed redesign \cite{10.1145/3491102.3517722}, and additional actions before users can continue consuming content \cite{ruiz2024design}. Across these settings, such interventions have reduced distraction and mindless scrolling \cite{lyngs2020just}, supported more intentional technology use and perceived agency \cite{haliburton2024longitudinal, 10.1145/3491102.3517722}, and improved attention or memory for consumed content, while also revealing recurring costs such as annoyance and frustration \cite{lyngs2020just, ruiz2024design}. Related interventions targeting information sharing similarly redirect attention toward considerations that frictionless interaction may obscure. Accuracy prompts, for example, can improve the quality of subsequently shared news \cite{pennycook2021shifting}, while proposed friction mechanisms for misinformation include reflection prompts, micro-exams, and other costs before sharing \cite{jahn2023friction}. Together, this literature positions friction not simply as a usability defect, but as a design resource for interrupting automatic action and redirecting users toward goals or information that might otherwise receive little attention.

In AI-assisted decision-making, cognitive forcing functions and output-level cues have been used to preserve independent judgment and encourage closer scrutiny of AI recommendations \cite{buccinca2021trust, gosline2024nudge}. In generative tasks, researchers have instead introduced friction into the production process itself: programming interfaces have required users to explain, predict, trace, or progressively reveal AI-generated code \cite{kazemitabaar2025exploring}; writing systems have structured feedback-driven revision around deliberate diagnosis, planning, and iteration \cite{zhang2025friction}; and prompt-entry interventions have encouraged writers to contribute more before generation by making short prompts slower to submit \cite{joshi2025writing}. Broader accounts of positive friction in human-AI interaction similarly argue that strategically slowing or complicating an interaction can create space for reflection rather than treating efficiency as the sole design objective \cite{chen2024exploring}. Across these systems, friction serves as a way of preserving human judgment, reflection, or contribution within AI-assisted work.

Prior work suggests that friction can support more thoughtful interaction with AI, but existing interventions are often coupled to a particular task or behaviour. We instead explore friction as a reusable layer of chatbot interaction, implemented at the level of individual conversational turns. Using writing as a task context, we compare six turn-level probes that vary how users contribute to, access, interpret, or use model responses, examining their effects on involvement, interaction costs, and workflow adaptation.

\section{Friction Design}
To guide our friction design, we conducted an exploratory synthesis of recent HCI work that deliberately interrupted otherwise smooth interaction. We reviewed 24 papers published within the past five years, identifying 73 unique designs and interventions. Most of these designs could be organized around three recurring interaction mechanisms: eliciting user contribution, restricting access, and reshaping system output. Guided by these mechanisms, we developed and iteratively refined a broader set of friction interfaces, ultimately retaining six design probes for evaluation.

\subsection{Deriving Friction Mechanisms from Prior HCI Work}
We used Google Scholar to assemble an exploratory corpus of 24 HCI papers published within the past five years, searching with combinations of terms related to friction and interaction regulation, including \textit{design friction}, \textit{self-control}, \textit{self-regulation}, \textit{constraint}, \textit{restriction}, and \textit{lockout}. Rather than limiting the corpus to AI chatbots, we considered designs across technology contexts, including digital wellbeing, social media, decision support, AI-assisted writing, and programming education. We retained papers containing concrete interaction interventions that introduced additional effort, delay, restriction, or other disruption into an otherwise familiar or smooth interaction. Because a single paper could contain multiple substantially different interventions, we treated each design as the unit of analysis. This resulted in a corpus of 73 unique designs from 24 papers. The full corpus is reported in \autoref{app:friction-corpus}.

The corpus served as a design resource for identifying recurring ways in which prior interfaces introduced friction. For each design, we examined what the intervention required users to do differently from the familiar interaction, for example, whether they had to contribute information, wait, inspect, interpret, verify, or transform content before proceeding. We iteratively compared these interaction patterns and grouped designs according to the primary way in which they altered the interaction. Most clustered around three recurring mechanisms: eliciting user contribution, restricting access, and reshaping system output. Four designs instantiated two mechanisms and were therefore coded under both, resulting in 77 design--mechanism entries in \autoref{app:friction-corpus}. Designs that did not clearly instantiate any of the three were retained under
\textit{Other}. Because these designs did not share a sufficiently coherent interaction pattern, we did not treat \textit{Other} as a fourth mechanism. We define the three mechanisms below.

\begin{enumerate} [label=M\arabic*., leftmargin=3.5em, labelsep= 0.8em, itemsep=0.6em]
\item
    \underline{\textit{Eliciting user contribution}} requires users to provide additional information, reasoning, judgment, or action beyond what the default interaction would ordinarily require.

    \item
    \underline{\textit{Restricting access}} delays, gates, or conditionally withholds content or functionality that would otherwise be immediately available.

    \item
    \underline{\textit{Reshaping system output}} changes the amount, form, organization, or communicative role of what the system presents, thereby changing how users must interpret or use it.
\end{enumerate}

\subsection{From Friction Mechanisms to Six Design Probes}
Guided by the three mechanisms, we developed and iteratively refined a broader set of design candidates before selecting six as probes. In adapting friction to AI chatbots, we distinguished our goal from that of many digital wellbeing and social media interventions, for which discouraging, interrupting, or terminating continued use may itself be an intended outcome. Our goal was not to discourage chatbot use as an end in itself, but to reshape how users engaged with AI assistance --- potentially reducing unreflective reliance while keeping users involved in the task and in decisions about how model responses were requested, interpreted, and used. We therefore considered friction productive when it redirected users' effort or attention toward the task, their own reasoning, or the model-generated content, rather than merely making the interaction more difficult or unpleasant.

Several early alternatives were not retained because they functioned primarily as obstruction. For example, we experimented with randomly alternating the capitalization of letters and adding shaking animations to model responses to slow reading, as well as replacing a proportion of nouns with symbols so that users would need to reconstruct their meaning. Although these alternatives reshaped system output, they impaired readability to the point that the responses became difficult to interpret and use. 

We also explored long-delay reveal designs that withheld a model response until a waiting period had elapsed.\footnote{This design was inspired by focus tools such as Forest \cite{forestappForestStay}, where a virtual tree must fully grow before a focus session is completed.} Such delays may be appropriate when the intended outcome is to discourage continued use or support disengagement from a distracting application. In a chat-based LLM exchange, however, an extended lockout appeared disproportionate to the short, iterative nature of individual turns and risked feeling punitive rather than supporting meaningful involvement~\cite{Mathur2021WhatMAA, gray2018dark}.

More broadly, we refined or discarded alternatives when they substantially overlapped with other candidates, were too easily bypassed, made model responses impractical to use, or directed users' effort primarily toward managing the interface rather than engaging with the task or the model-generated content. From this broader set, we selected six design probes for further evaluation. Each mechanism was represented by two distinct probes, allowing us to examine variation across different implementations of the same mechanism. At the same time, each probe was designed around a single primary mechanism to support clearer comparative interpretation. 

\subsection{Friction Prototypes}
We organized the six probes according to the three friction mechanisms. 

\subsubsection{\underline{Eliciting User Contribution}} 
\mbox{}\par

\begin{figure*}[h]
	\centering
	\includegraphics[width=\twocolwidth]{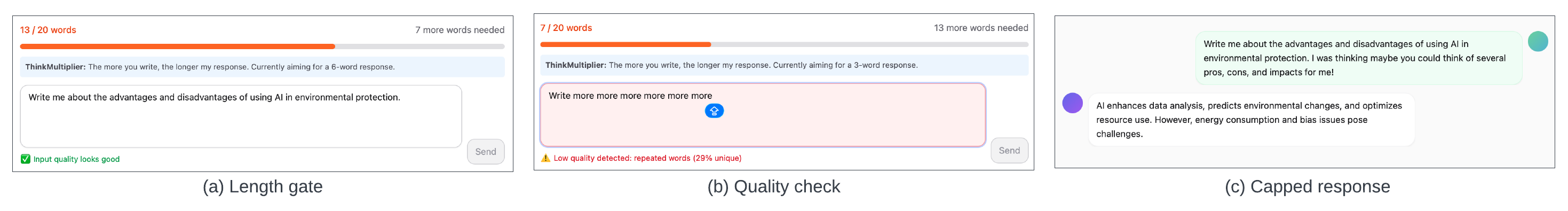}
    \caption{The Think Multiplier interface and its input and output constraints. (a) A 13-word prompt falls below the 20-word minimum, so the probe blocks submission. (b) Repeated words do not count toward the minimum, preventing users from padding the prompt. (c) The generated response is capped at around half the length of the user’s prompt.}
	\label{fig:think_multiplier} %
\end{figure*}

\textbf{\textsc{Think Multiplier}} ---  
Before receiving a response, users must compose a prompt of at least 20 words (\autoref{fig:think_multiplier}a). We introduced this threshold to discourage short, search-like prompting (often fewer than 15 words) \cite{trippas2024users} and instead require some articulation of context and intent. During refinement, we found that a length requirement alone could be easily bypassed, so we enforced a prompt quality check that rejects prompts containing the same sentence repeated more than three times (\autoref{fig:think_multiplier}b). The model’s reply is also capped at around half the length of the user’s input (\autoref{fig:think_multiplier}c), ensuring that users remain the primary authors while the AI plays a subordinate role.  

\smallskip

\textbf{\textsc{User Echo}} --- After submitting a prompt, users must write their own tentative response before the model replies (\autoref{fig:teaser}b). The model first provides constructive feedback on the user's response and then builds on it with additional ideas, perspectives, or information. By requiring users to formulate an initial response before seeing the model output, \textsc{User Echo} was intended to preserve their involvement in planning and idea generation, cognitive processes that prior work has associated with greater psychological ownership \cite{reza2025co}.

\subsubsection{\underline{Restricting Access}} 
\mbox{}\par

\textbf{\textsc{Hold Reveal}} --- 
The model response is blurred and becomes visible only while users simultaneously hold down keys 1, 2, 9, and 0, located near the corners of the keyboard (\autoref{fig:teaser}c). We deliberately chose a key combination that occupies both hands, requiring users to maintain a continuous physical action to access the response while making it impractical to select
or copy the text at the same time. This embodied access control draws on interaction work that links physical effort to attention and sense-making \cite{antle2011embodied}. 

\smallskip

\begin{figure*}[h]
	\centering
	\includegraphics[width=\twocolwidth]{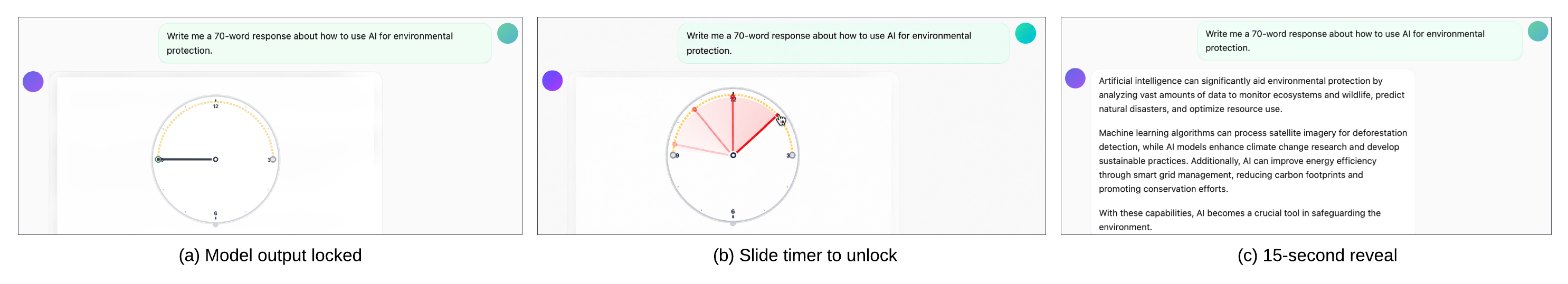}
    \caption{Timer Lockout. (a) Model output is initially hidden. (b) Users slide the timer to unlock the output. (c) The output is revealed for 15 seconds before automatically re-locking.}
	\label{fig:timer_lockout} %
\end{figure*}

\textbf{\textsc{Timer Lockout}} --- 
Model output is hidden behind an overlay and only revealed in 15-second intervals when users slide a timer (\autoref{fig:timer_lockout}); once the interval ends, the text re-locks. We chose a 15-second window to introduce lightweight temporal friction: it is sufficient to read and form a gist of a short passage, while preventing prolonged, unreflective skimming. Inspired by prior work using timed lockouts to disrupt habitual access and encourage re-engagement \cite{haliburton2024longitudinal}, this probe segments access to model text. After initial testing, we also disabled copy-paste so that users could not move the model output to their writing textbox (either to read it without the lock-out or to use it verbatim in their response). 

\subsubsection{\underline{Reshaping System Output}} 
\mbox{}\par

\textbf{\textsc{Less is More}} --- 
This interface limits output to one sentence per turn (\autoref{fig:teaser}e). Drawing on Minimalism \cite{redish1998minimalism} and Cognitive Load Theory \cite{sweller2011cognitive}, we treated concise output as a way to encourage active meaning-making. Prior studies also shows that sentence-level suggestions can preserve user agency and reduce automation bias more effectively than longer, document-level responses \cite{fu2023comparing}. 

\smallskip

\textbf{\textsc{Elenchus}} --- 
Inspired by Socratic dialogue \cite{knezic2010socratic}, this design refuses to provide answers. Instead, the LLM responds with questions, such as "Have you considered...?", that challenge the user's premises (\autoref{fig:teaser}f). 

\subsection{Implementation}
The interface is built with React (Vite) and a Node.js backend. We use GPT-4o via the OpenAI API with in-context learning. Each friction condition is implemented as a modular React component. During usage, the app logged user interactions for analysis.

\section{Experimental Method}
To evaluate the six design probes, we conducted a within-subjects study in which participants completed multiple short writing tasks and a brief survey after each task. We selected writing as the task context because it is a common use case for AI chatbots \cite{NBERw34255} and allowed us to examine how users requested, interpreted, and incorporated model-generated content within a bounded task. Because our primary interest was turn-level chatbot interaction rather than extended writing processes, we prioritized breadth across friction designs and therefore used short, bounded tasks. After completing the writing tasks, participants took part in a semi-structured interview. They then completed recall and recognition tests, which complemented the post-task self-report measures by assessing their memory for their own writing. 
Specifically, we aimed to answer the following questions:
\begin{enumerate} [label=RQ\arabic*:, leftmargin=3.5em]
    \item How do different design frictions affect users' involvement when working with an AI chatbot?
    \item What interaction costs do different design frictions introduce?
    \item How do users adapt their workflows in response to design frictions?
\end{enumerate}

\subsection{Participants}
We recruited 24 participants (age range: 18 - 44; 12 self-identified as male, 12 as female). 23 participants reported prior experience with generative AI tools for writing, including ChatGPT (22), Gemini (15), Claude (11), and DeepSeek (7). On a 5-point Likert scale from 1 = strongly disagree to 5 = strongly agree, participants reported they relied on LLMs for their writing tasks (Mdn = 3, SD = 1.36) and felt a lack of control over their writing when using these tools (Mdn = 4, SD = 0.73). Weekly usage of generative AI for writing varied widely (M = 5.33, SD = 9.22; range: 0 - 40 hours). 

\subsection{Apparatus}
We conducted the study either in person or online. In-person sessions were held in a quiet lab environment, where each participant used a provided laptop equipped with a web browser to access the experimental interface. Remote sessions were conducted via Microsoft Teams. All sessions were audio and screen recorded with participants' consent, and the study software automatically logged all on-screen interactions, including writing responses, human-LLM exchanges, and time spent on each task. Each session lasted approximately an hour. Participants received a \$30 Amazon gift card as compensation for their time and effort. The study protocol was reviewed and approved by the relevant institutional research ethics board. 

\subsection{Procedure, Design, and Tasks}
The study followed a within-subject design, where each participant experienced all eight interface conditions: six frictional LLM interfaces and two non-frictional conditions (ChatGPT and Write by Yourself). While ChatGPT served as a baseline, Write by Yourself provided a non-LLM control condition that allowed us to observe how participants write without any model assistance. To mitigate order effects, these conditions were counterbalanced using a balanced Latin square. Participants completed four sequential tasks: writing, memory refresh, recall, and recognition, and we inserted a brief semi-structured interview between the writing and memory-refresh phases. We selected this placement because participants had just engaged with the frictional interfaces during the writing task, allowing them to articulate their experiences while those interactions were still fresh in mind. 

\medbreak
\textit{Introduction (\textasciitilde{}2 mins)} - Participants first completed an online consent form and a demographics survey before beginning the experimental tasks.

\begin{figure*}[tb]
	\centering
	\includegraphics[width=0.8\twocolwidth]{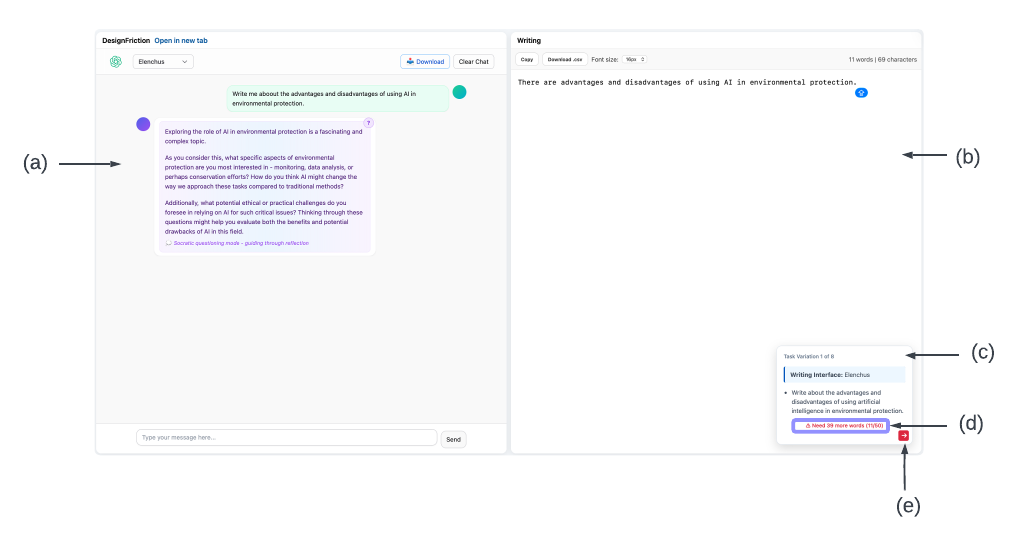}
	\caption{Task 1 study interface: (a) the LLM interaction area; (b) the writing board; (c) the writing topic display; (d) the real-time word-count check that must reach the minimum requirement before proceeding; (e) the button used to advance to the evaluation page.}
	\label{fig:task1} %
\end{figure*}

\medbreak
\textit{Tasks 1 (Writing; \textasciitilde{}30mins)} - Each participant completed eight writing trails (8 interfaces $\times$ [1 writing sub-task + 1 survey evaluation]). During each sub-task, they were instructed to make use of the LLM-supported interface for that condition to complete a short writing prompt. Responses were required to be between 50 and 80 words and to form coherent, grammatically complete paragraphs; participants could not proceed to the survey evaluation page until the word-count requirement was met (\autoref{fig:task1}d). The task interface consisted of two panels: the LLM interaction area on the left (\autoref{fig:task1}a) and the writing board on the right (\autoref{fig:task1}b). The writing topic appeared in the bottom-right corner of the screen (\autoref{fig:task1}c).
We selected prompts adapted from IELTS-style writing tasks, which do not require external research or factual knowledge, allowing participants to rely solely on the LLM-assisted interface. Topics spanned diverse domains, such as technology, agriculture, government, to minimize topic-familiarity bias. A full list of writing prompts can be found in \autoref{apx:prompts}.

For each sub-task, participants interacted with one of the six frictional interfaces or a baseline condition, composed their response in English, and then proceeded to an evaluation page by clicking the ``$\rightarrow$'' button (\autoref{fig:task1}e). On this page, they rated:

\begin{itemize}
    \item Workload using NASA-TLX subscales (mental demanding, physical demanding, performance, effort, and frustration) \cite{hart1988development}.
    \item Personal ownership on a 5-point scale (1 = strongly disagree, 5 = strongly agree) in response to the statement "I feel a strong sense of personal ownership over this writing."
\end{itemize}

After completing the ratings, participants advanced to the next interface condition. 

\medbreak
\textit{Semi-structured Interview (\textasciitilde{}10 mins)} - After completing eight writing trails, a semi-structured interview was conducted with the experimenter. Questions were open-ended (see \autoref{apx:questions} for full list) and explored participants' overall experiences with the different design frictions, their perceived difficulty and ease of use, and their impressions of how each design influenced their sense of ownership and control during writing.

\medbreak
\textit{Task 2 (Memory Refresh; \textasciitilde{}5 mins)} - Once interview was finished, participants performed 50 elementary-level arithmetic problems, such as 3 $\times$ 7 = and 18 - 6 = . This task aimed to disrupt short-term memory, which holds recently encountered information for brief periods \cite{cowan2008differences, clevelandclinicWhatShortTerm}, and to reduce recency effects, where the most recent content is recalled more easily than earlier material \cite{baddeley2022recency}. Prior cognitive psychology work shows that brief interference tasks, typically 15–30 seconds, are sufficient to attenuate these effects \cite{mcleod2024serial}. Based on this evidence, we used a simple arithmetic task as the interference activity before the recall phase, and no correctness feedback was provided to avoid stress or performance pressure. 

\medbreak
\textit{Task 3 (Recall; \textasciitilde{}15 mins)} - Participants were then asked to reproduce as much of their original writing from Task 1 as possible. To help orient memory, the eight writing topics were shown as cues. Participants were encouraged to restate their central arguments, phrasing, and supporting examples from memory. To complement self-reported involvement, we assessed memory as a behavioural indicator of participants' cognitive involvement in the writing task and the extent to which they internalized the text they produced. Prior work has similarly linked recall of one's own AI-assisted writing to memory encoding and cognitive engagement \cite{kosmyna2025your}. 

\medbreak
\textit{Task 4 (Recognition; \textasciitilde{}5 mins)} – The final task was a sentence-recognition test, designed to complement the recall task by capturing memory for finer-grained textual details. We prepared eight recognition questions corresponding to the eight writing sub-tasks. For each question, participants were asked to identify the exact sentence they had written in Task 1 from four options. One option was a sentence randomly drawn from the participant’s own writing. The remaining three foils were generated using GPT-4o: a paraphrased version that preserved the meaning but altered the wording, an opposite version that inverted the original meaning, and an unrelated sentence that matched the length but addressed a different topic. This design allowed us to examine whether participants remembered only the general topic, the semantic gist, or the precise sentence they had produced.

\subsection{Data Analysis}
Quantitative differences across writing interfaces were examined using the Friedman test. When the omnibus test was significant, we conducted pairwise Wilcoxon signed-rank tests with 10,000 replications and applied Holm correction to account for multiple comparisons. All reported differences are computed as interface – baseline (i.e., positive values indicate higher values for the interface). 

We analyzed the interview transcripts and interaction logs (prompt–response histories) using inductive thematic analysis \cite{braun2021thematic}. The first author generated initial codes grounded in the data, focusing on participants’ comments and recurring behavioural patterns in how they engaged with, adapted to, or resisted different frictions. The second author then reviewed these codes with the first author, and the coding scheme was iteratively refined through discussion and repeated passes over the dataset. Finally, the refined codes were collated into higher-level themes that described common patterns in participants’ perceptions of the frictions and the workflow adaptations they adopted across conditions.  

To evaluate memory outcomes, we designed a recall scoring pipeline. The first author de-identified all participant responses by removing names and interface labels. Each response was then segmented into three to five meaning-balanced sentences of roughly equal length. This ensured that each segment carried a comparable amount of information, avoiding unfair penalization of responses that were longer or more syntactically complex.
Each sentence–recall pair was independently scored by two researchers using a three-point rubric: 0 for no recall, 0.5 for partial recall, and 1 for full recall. Scoring examples are provided in \autoref{tab:semantic-scoring}. After scoring, the first author identified all disagreements and facilitated group discussion among all researchers to reach consensus on final scores. A participant's recall score for each interface was computed as: 
\[
\text{Recall}_{i} = \frac{1}{N} \sum_{j=1}^{N} s_{ij}
\]
where \( N \) is the total number of segments, and \( s_{ij} \) is the recall score assigned to the \( j \)-th segment in interface \( i \).

\section{Results}
All participants completed the study. However, due to a keyboard technical issue, P3 was unable to complete the task associated with the \textsc{Hold Reveal} interface. As a result, analyses involving \textsc{Hold Reveal} are based on data from 23 participants, while all other interface conditions are evaluated with data from the full sample of 24 participants. 

\subsection{Workload}
Participants reported significantly higher overall workload ($p \leq .007$) across all friction conditions compared to the ChatGPT baseline (\autoref{fig:workload}a); subscale-level results can be found in \autoref{app:workload}. Following established practice \cite{devos2020psychometric}, overall workload was calculated as the unweighted mean of the five NASA-TLX subscales. 
Notably, many friction conditions increased workload over the \textsc{Write by Yourself} condition.

At the subscale level, all friction conditions significantly increased \textit{Mental Demand} ($p \leq .002$) and \textit{Effort} ($p < .001$), and frustration relative to ChatGPT. \textit{Physical Demand} was also significantly higher for most conditions ($p \leq .011$), with \textsc{User Echo} as the exception ($p = .06$), while no significant differences were found for \textit{Performance}.

\subsubsection{Added Cognitive Effort Could Support or Interfere with Thinking.} 
For some, the additional effort was directed toward the writing itself. P4 felt that \textsc{Think Multiplier} \textit{``enforced me to think a lot and write a lot of things,''} but also noted that this could produce \textit{``better answers''} in which \textit{``most points would come from my thoughts.''} P13 similarly described \textsc{Elenchus} as \textit{``pretty helpful''} because its questions were \textit{``giving directions on what you can think about,''} particularly when they wanted to engage in \textit{``critical thinking''} and write in their own words.

For others, the additional effort interfered with the task. P3 felt that some interfaces \textit{``hinder[ed] my reasoning''} and required \textit{``extra effort in order to achieve the same level of performance''} and obtain the same amount of information. P13 similarly characterized the repeated loss of access in \textsc{Timer Lockout} and \textsc{Hold Reveal} as \textit{``more cognitive overload,''} explaining that they became occupied with \textit{``how to deal with this sudden loss of info''} rather than developing their ideas.

\subsubsection{Physical Demand was Most Pronounced for the Access-restricting Interfaces.}
\textsc{Timer Lockout} and \textsc{Hold Reveal} increased physical demand by 5.63 and 5.69, respectively, relative to ChatGPT. Participants connected this demand to the embodied actions required to access model output. P9, for example, noted that \textsc{Hold Reveal} occupied both hands, such that they \textit{``couldn't even scroll through the text properly.''}

\subsubsection{Frustration Emerged When Friction Felt Obstructive or Disproportionate}
\textsc{Think Multiplier} produced the largest increase relative to ChatGPT ($\Delta M=7.58$), closely followed by \textsc{Timer Lockout} ($\Delta M=7.54$), whereas \textsc{User Echo} showed the smallest increase ($\Delta M=2.29$). Participants attributed their frustration to different aspects of the friction. For
\textsc{Think Multiplier}, several participants questioned whether the effort required was worthwhile given the assistance returned. P13 described having to \textit{``supply 100 words to get back 50 words''} as \textit{``not a good trade-off,''} while P23 felt that they were writing \textit{``things for the sake of writing things.''} \textsc{Timer Lockout} was frustrating in a different way because it repeatedly interrupted access to an already generated response. P10 felt that \textit{``the only use for that would be to frustrate''} the user, while P12 described losing access before finishing reading and having to reveal the response again as \textit{``a waste of time.''} Frustration with \textsc{Elenchus} was more often tied to the kind of assistance it provided. P6, for example, already knew the direction of their argument and wanted \textit{``information or things to back up my thoughts,''} making the model's continued questioning frustrating rather than helpful. 

\begin{figure*}
\centering
\includegraphics[width=\twocolwidth]{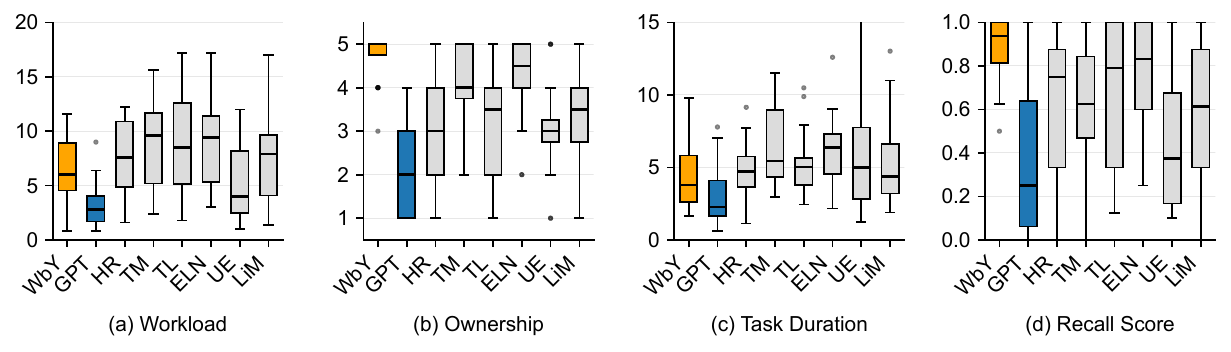}
\caption{Boxplots show the distribution of (a) NASA-TLX scores by interface, (b) ownership ratings on a 5-point Likert scale, (c) task duration (mins), and (d) recall score (0–1). Abbreviations for conditions: \textsc{Write by Yourself} (WbY), \textsc{ChatGPT} (GPT), \textsc{Hold Reveal} (HR), \textsc{Think Multiplier} (TM), \textsc{Timer Lockout} (TL), \textsc{Elenchus} (ELN), \textsc{User Echo} (UE), and \textsc{Less is More} (LiM). We highlight \textsc{Write by Yourself} in yellow and \textsc{ChatGPT} conditions in blue; all friction conditions are in grey.}
\label{fig:workload} 
\end{figure*}

\subsection{Ownership}
\label{res:ownership}
All friction interfaces, as well as the \textsc{Write by Yourself} condition, significantly increased participants’ sense of ownership compared to ChatGPT ($M = 2.00$); see \autoref{fig:workload}b. On a 5-point scale, \textsc{Write by Yourself} increased ownership ratings by 2.71, \textsc{Elenchus} by 2.25, \textsc{Think Multiplier} by 1.92, \textsc{Less is More} by 1.38, \textsc{Timer Lockout} by 1.21, \textsc{User Echo} by 1.04, and \textsc{Hold Reveal} by 1.00. Unlike the workload measure, where the \textsc{Write by Yourself} condition did not have the highest workload, ownership was highest for the \textsc{Write by Yourself} condition.

\subsubsection{Definition of Ownership}
Participants described ownership as extending beyond who typed the final words to include whose ideas, expression, and thinking shaped the response, as shown in \autoref{tab:ownership-definition}. 
\begin{table*}[h]
\centering
\caption{Participants' definitions of ownership in AI-assisted writing.}
\label{tab:ownership-definition}
\small
\begin{tabularx}{\textwidth}{
    @{}
    >{\raggedright\arraybackslash}p{0.45\textwidth}
    >{\raggedright\arraybackslash}X
    >{\centering\arraybackslash}p{0.035\textwidth}
    @{}
}
\toprule
\textbf{Definition} &
\textbf{Representative quote} &
\textbf{$n$} \\
\midrule

Ownership increases with the human share of the work and decreases with
reliance on AI-generated content. &
``It depends how much assistance I get from the LLM \ldots\
[if it] barely involved my thinking \ldots\ less ownership.'' (P16) &
17 \\
\addlinespace

The ideas, positions, perspectives, or substantive content originate from
the user. &
``If I feel like the idea belongs to me, I have a sense of ownership over
the overall answer.'' (P12) &
14 \\
\addlinespace

The user contributes wording, sentence structure, style, organization, or
narrative flow. &
``Every time I felt like I was \ldots\ coming up with my own sentence
structures \ldots\ I really felt more ownership.'' (P1) &
12 \\
\addlinespace

Ownership comes from thinking, reflecting, synthesizing, justifying, or
working through how the response is developed. &
``I've put my brain power into it \ldots\ I have taken information and I
have synthesized.'' (P14) &
11 \\
\addlinespace

The user intentionally steers, selects, manipulates, or shapes the final
output. &
``[Ownership is] to what extent I can manipulate \ldots\ the final
response.'' (P3) &
5 \\
\addlinespace

The user understands and identifies with the text and can explain or defend
what was written. &
``If you quiz me on whatever I've written, I know exactly what answer to
give \ldots\ then it's my own.'' (P13) &
5 \\

\bottomrule
\end{tabularx}
\end{table*}

\subsubsection{What Shaped Ownership Across Conditions}
The larger mean increases in ownership observed for \textsc{Write by Yourself} and \textsc{Elenchus} were broadly consistent with participants' emphasis on their own contribution to the writing. Participants frequently described \textsc{Elenchus} as requiring them to develop the substance of the response rather than receiving ready-made prose. P19, for example, explained that \textit{``you do almost all the reflection yourself.''} \textsc{Think Multiplier} and \textsc{Less is More} were described similarly as limiting how much finished text the model supplied, leaving participants to develop more of the response themselves. 

The retrospective interviews, however, did not mirror the quantitative pattern exactly. \textsc{User Echo} showed a comparatively small ownership increase, yet was the second most frequently named friction when participants were later asked which interface gave them the greatest sense of ownership. Participants often valued that it elicited their own position before the model elaborated on it. P21, for example, appreciated that it \textit{``actually asked me about [...] what's your idea on this before it writes me a response.''} For these participants, ownership could come from preserving the origin and direction of an idea even when the model contributed to its eventual expression.

The access-restricting interfaces reflected another route to ownership. Participants described \textsc{Timer Lockout} and \textsc{Hold Reveal} as making model output difficult to reuse directly, requiring them to reconstruct or write the response themselves. As P16 put it when describing \textsc{Timer Lockout}, \textit{``I just wrote it myself.''} Here, increased ownership could arise from reduced reliance on the model's prose rather than from additional support for developing the user's ideas.

\subsection{Task Duration}
All friction interfaces required more time to complete the writing task than the baseline ($p<.001$).
Compared to ChatGPT's average of 3.15 minutes, \textsc{Think Multiplier} showed the largest increase (3.41 mins), followed by \textsc{Elenchus} (2.96 mins). \textsc{User Echo}, \textsc{Less is More}, \textsc{Timer Lockout}, and \textsc{Hold Reveal} formed a middle range with broadly similar completion times. Interestingly, all frictions had a higher average task duration than the \textsc{Write by Yourself} condition (\autoref{fig:workload}c).

\subsection{Recall Performance}
\label{res:recall}
Participants were asked to recall in a free text box as much as they could of what they wrote, after doing a distracter task; recall performance was graded on a 0 (no recall) to 1 (perfect recall) scale (\autoref{fig:workload}d). \textsc{Write by Yourself} had the highest average recall score ($M = 0.88$), and most interfaces improved participants’ ability to recall their written content compared to ChatGPT ($M = 0.38$), with the exception of \textsc{User Echo}. The largest recall gain among the frictions was observed for \textsc{Elenchus} ($\Delta M = 0.40$, $p =.002$).

\subsection{Recognition Performance}
\label{res:recognition}
Participants were also asked to select the sentence they had written from a set of distractor sentences. Overall participants did quite well on this task. Participants in the ChatGPT condition had an average recognition accuracy of 0.63, with only \textsc{Less is More} ($\Delta M = 0.38$, 95\% CI [0.17, 0.59], $p =.02$) and \textsc{Elenchus} ($\Delta M = 0.33$, 95\% CI [0.14, 0.55], $p =.04$) seeing improved recognition performance.
No significant differences were observed for the remaining conditions, perhaps due to the task being easier than the free recall task, although all conditions had higher averages than ChatGPT (\autoref{app:workload}).

\subsection{Workflow Adaption Patterns}
Interaction logs showed that participants adapted to the frictions by changing how much they interacted with the model and how they distributed writing work across turns. We observed four recurring adaptations.

\subsubsection{Breaking the Writing Task into Smaller Steps}

Participants often adapted by turning a paragraph-level writing request into a sequence of smaller questions or writing stages. This was particularly visible with \textsc{Less is More}, where a single-sentence response could become a starting point for the next request rather than a finished answer. For example, after initially asking whether nuclear power should meet future energy needs, P12 separately asked for \textit{``the advantages,''} \textit{``the disadvantages,''} ways to \textit{``overcome the disadvantages,''} and why nuclear power was important for the future. P18 decomposed the writing process even more explicitly, moving from \textit{``give me a thesis statement''} to \textit{``the first claim,''} \textit{``the explanation of that,''} \textit{``evidence,''} and finally \textit{``specific examples.''} Other participants progressively narrowed broad questions into particular mechanisms, examples, or counterarguments. This was also reflected in the number of prompts participants sent:
\textsc{Less is More} averaged 6.04 turns ($SD=3.62$), while
\textsc{User Echo} ($M=3.46$, $SD=2.54$), \textsc{Elenchus}
($M=3.42$, $SD=2.21$), and \textsc{Think Multiplier}
($M=3.38$, $SD=1.69$) also involved more turns on average than ChatGPT ($M=2.13$, $SD=1.30$).

\subsubsection{Bringing User Ideas Earlier into the Interaction}

Another adaptation was to place participants' own ideas or drafts earlier in the interaction and use subsequent turns to develop them. This pattern was especially visible in \textsc{User Echo}, where participants often responded to the required intermediate turn with a substantive position rather than only a minimal answer. P11, for instance, first supplied their own knowledge about the technological benefits of space exploration and its relative government spending, and then asked the model to \textit{``summarize that in 50 to 75 words''} using \textit{``my words and some additional insight provided by you.''} Similar co-development appeared in other interfaces when participants introduced their own stance and subsequently asked the model to elaborate, check, or integrate it. Rather than moving directly from a question to model-generated prose, these workflows introduced an intermediate artifact from the participant that later model output was built around.

\subsubsection{Compressing Model Output to Fit Access Constraints}

Participants also changed the granularity of the information they requested when model output was difficult to retain or revisit. With \textsc{Hold Reveal}, for example, P10 first requested advantages of genetic modification, then asked the model to \textit{``summarize this into a paragraph''} and subsequently to \textit{``reduce this to a couple sentences''} before moving on to the disadvantages. P24 similarly reformulated a request for arguments about nuclear power as \textit{``a short set of bullets.''} Under \textsc{Timer Lockout}, P12 broke the requested information into individual sentences --- one summarizing genetic modification, one describing its benefits, and one its risks. Here, participants adapted less by changing the substance of the writing task than by making model output smaller and easier to capture before access was restricted again.

\subsubsection{Working Around Friction to Restore Familiar Workflows}
Not all workflow adaptations redirected effort toward the writing itself. Participants also developed strategies for satisfying or bypassing a constraint while attempting to recover more conventional chatbot behaviour. With \textsc{Think Multiplier}, P18 appended long strings of repeated characters (e.g., \textit{``a a a \ldots b b b \ldots''}) to otherwise ordinary requests, while P3 repeatedly inserted previous model-generated text into subsequent prompts. In these cases, the additional input increased prompt length without necessarily introducing new reasoning.

\section{Discussion}
Our findings paint a mixed picture of design friction in AI chatbot interactions. Regarding user involvement (RQ1), all six friction probes increased perceived ownership relative to ChatGPT, while their effects on memory were more selective: most improved recall, but only \textsc{Elenchus} and \textsc{Less is More} significantly improved recognition of participants' own wording. Regarding interaction costs (RQ2), these benefits came at a consistent cost, as every friction increased workload and task duration. Finally, participants did not simply comply with the constraints (RQ3); they adapted by decomposing tasks, bringing their own ideas earlier into the interaction, compressing model responses, or working around the friction to restore familiar chatbot workflows. These results shift the question from whether friction can increase involvement to when and how it can do so productively.

\subsubsection*{First, there was no universally productive friction: the same constraint could support or obstruct users depending on when it appeared and what they wanted from the AI} This was particularly visible with \textsc{Elenchus}. P6, for example, had already decided the direction of their argument and wanted the model to provide \textit{``information or things to back up my thoughts''}; continued Socratic questioning therefore interfered with rather than supported the assistance they were seeking. Other participants described workflows in which they preferred to develop their writing themselves and turn to AI later for editing or grammatical support. In these cases, a friction designed to elicit further ideation may reproduce work the user has already done. Conversely, when users have not yet formed a position, the same questioning interaction may prevent them from immediately outsourcing that reasoning to the model. This tension echoes prior work on cognitive forcing functions, which found that interventions that reduce overreliance can nevertheless be less preferred by users \cite{buccinca2021trust}. Our findings suggest that one reason for this tension is fit: friction preserves useful human work only when that work is still relevant to the user's current goal. Rather than asking which friction should be applied universally, designers may therefore need to consider when friction is appropriate, for example, whether users are forming an argument, seeking supporting information, drafting, or revising.

This also raises a question about our decision to scope friction at the turn level. Turn-level interventions offer a lightweight and modular design unit, but applying the same friction across turns can overlook how users' goals change over the course of a task. This tension was also visible when participants were asked to design their own model. Rather than converging on a single interaction style, many described systems that could switch between reflective questions and direct assistance, or allow particular forms of friction to be turned on and off depending on the task. P6, for example, wanted a model that could guide with questions during ideation but provide specific examples when directly requested, while P24 considered Socratic questioning useful for studying but wanted it to be optional for writing. This tension connects to mixed-initiative interaction, which emphasizes that system intervention should account for users' goals and the timing and costs of assistance rather than treating initiative as fixed \cite{796083}. Our findings therefore suggest that turn-level friction may be more useful as a repertoire of interventions that users can invoke as needed than as a uniform rule applied throughout an interaction. 

\subsubsection*{Second, our memory results suggest that productive friction needs to preserve meaningful human work beyond a single moment of participation.} The contrast between \textsc{User Echo} and \textsc{Think Multiplier} is revealing because both elicited greater user contribution, yet only \textsc{Think Multiplier} significantly improved recall over ChatGPT; \textsc{User Echo} was the only friction for which the recall difference was not statistically significant. \textsc{User Echo} clearly succeeded at bringing participants' ideas earlier into the interaction. Participants often supplied a substantive position before the model responded and then used the AI to elaborate or rewrite it. Yet once this initial contribution had been made, the model could still provide relatively complete prose that participants could directly adopt. One possible explanation is that contributing an idea is only one stage of involvement: it does not necessarily ensure that users carefully read, evaluate, or integrate what the model subsequently produces. This possibility is consistent with prior work showing that people can readily rely on LLM-generated responses, including generations presented without explicit expressions of confidence~\cite{rathi2025humansoverrelyoverconfidentlanguage}. \textsc{Think Multiplier}, in contrast, maintained a proportional constraint on the division of work: users were required to contribute more text than the model could return, keeping the user as the larger contributor throughout the exchange. We cannot attribute the recall difference to this property alone, since the two probes differed in several ways. However, their contrast suggests that requiring an initial contribution may not be enough; friction may also need to preserve user involvement when model-generated content is subsequently produced and incorporated into the final response.

The recognition results reinforce this interpretation from another angle. The recognition task required participants to distinguish the exact sentence they had written from alternatives including a paraphrase with the same meaning, making it a relatively fine-grained test of memory for their own wording. Only \textsc{Elenchus} and \textsc{Less is More} significantly improved recognition over ChatGPT. Importantly, these were also the two probes that most directly prevented the model from supplying a ready-to-adopt response. \textsc{Elenchus} provided questions rather than prose, requiring participants to translate its prompts into their own argument and wording. \textsc{Less is More} still provided content, but only one sentence at a time, and participants often responded by progressively constructing their answer through a thesis, claims, explanations, evidence, and examples; it consequently involved substantially more conversational turns than ChatGPT. This pattern resonates with the generation effect, in which self-generated information is remembered better than material that is simply presented \cite{slamecka1978generation}, and with the ICAP framework's distinction between passively receiving information and constructively generating beyond what has been provided \cite{Chi02102014}. We do not claim that our recognition task directly tests either theory, but they provide a lens for interpreting the difference: \textsc{Elenchus} and \textsc{Less is More} left participants responsible for more of the formulation and assembly of the final response. Broadly speaking, for AI chatbot uses where understanding matters, such as learning or debugging, this suggests that the goal of friction may not be to make an answer harder to obtain, but to make the reasoning needed to construct or understand that answer harder to bypass. Related work in programming education similarly explores requiring learners to predict, explain, trace, or reason about AI-generated code rather than simply receiving complete solutions \cite{kazemitabaar2025exploring}.

\subsubsection*{Third, restricting-access friction appeared to influence AI reliance through a different route: by changing when users considered AI assistance worth seeking or reusing.} Unlike the other probes, \textsc{Hold Reveal} and \textsc{Timer Lockout} did not lead participants to send more prompts than they did with ChatGPT. Participants instead often adapted the size of the assistance they requested. Under \textsc{Hold Reveal}, for example, P10 progressively asked the model to reduce information to a paragraph and then \textit{``a couple sentences,''} while P24 requested \textit{``a short set of bullets.''} Under \textsc{Timer Lockout}, P12 broke the information they needed into individual sentences. Others reduced their reliance on the generated response altogether; P16 summarized their experience with \textsc{Timer Lockout} simply as, \textit{``I just wrote it myself.''} Rather than encouraging more iterative collaboration with the model, access friction appeared to raise the cost of repeatedly seeking, revisiting, or reusing AI assistance. This interpretation connects to cost-benefit accounts of AI reliance, which argue that people strategically weigh the effort required to engage with a task against the expected benefit of relying on AI \cite{10.1145/3579605}. \citet{10.1145/3579605} for example, showed that changing the costs of engaging with AI information can change reliance behaviour. Restricting access may similarly make an additional AI response no longer feel effectively free, encouraging users to decide whether the assistance is worth requesting. However, reduced help-seeking should not be equated with deeper engagement. Participants also described \textsc{Timer Lockout} and \textsc{Hold Reveal} as cognitively or physically distracting: some became occupied with maintaining or recovering access rather than developing their ideas. This points to an important distinction between different goals for friction. If the goal is to discourage habitual or unnecessary reliance, increasing the cost of seeking AI assistance may itself be useful. If the goal is instead to improve how users understand and work with information they genuinely need, our findings suggest that friction may be better placed in the acts of interpreting, formulating, or assembling the response rather than around access alone.

\section{Limitations and Future Work}
\subsubsection*{Writing tasks were short and bounded.}
Our goal was to study turn-level friction in AI chatbot interactions. We used short writing tasks because they provided a bounded setting in which participants could request, interpret, and incorporate model output under each friction condition, allowing us to evaluate multiple friction probes. However, our findings also show why these results should not be interpreted as identifying universally effective or ineffective frictions: the value of a friction depended on what users were trying to accomplish and where it intersected with their workflow. Even within these short writing tasks, we saw participants note that in different stages they used the AI chatbot differently and had different expectations for support. A friction that is productive during ideation, for example, may be obstructive during information seeking.  Future work should evaluate turn-level friction in longer tasks contexts, where users are more likely to use chatbots in a more diverse set of ways.

\subsubsection*{Static rather than adaptive friction.} The current prototypes applied friction in fixed ways, regardless of users’ intentions or changing needs during the task. This rigidity sometimes turned potentially constructive friction into frustration, especially when participants wanted to move between deeper engagement and efficient information access. Future work could explore more legible and adjustable forms of friction, including lighter-weight nudges or systems that allow users to change constraints over time. Our Usage Mirror prototype (\autoref{tab:usage_mirror}) points to one such direction by foregrounding self-monitoring.

\subsubsection*{Optional frictions and user workarounds.} Most frictions can be avoided if the user is motivated enough to turn them off. While one avenue is to seriously increase the level of friction---like an alarm clock that runs away from you or requires you to complete math problems to turn off the alarm---another is to design frictions that users are more likely to voluntarily leave on. In July, 2025 ChatGPT introduced Study Mode,\footnote{\url{https://openai.com/index/chatgpt-study-mode/}} ``a learning experience that helps you work through problems step by step instead of just getting an answer.'' This suggests there is interest in certain kinds of frictions that can be turned on and off by users at will. Future work could consider frictions that can be turned on and off by users, and investigate when and why users activate them over the course or days or weeks.

\section{Conclusion}
We presented six turn-level friction probes for AI chatbot interactions, using writing as a task context to study how friction affects user involvement, interaction costs, and workflow adaptation. A within-subject study with 24 participants showed that friction consistently increased perceived ownership, workload, and task duration, while its effects on memory varied across designs. Participants also adapted their workflows in response to different forms of friction. Broadly, our work suggests that productive friction depends not on making AI interaction harder, but on preserving meaningful human work when it matters.

\begin{acks}
\end{acks}

\bibliographystyle{lib-acm/ACM-Reference-Format}
\bibliography{_references.bib}

\appendix
\makeatother
\clearpage
\renewcommand\thefigure{\thesection.\arabic{figure}}
\renewcommand\thetable{\thesection.\arabic{table}}
\setcounter{figure}{0}
\setcounter{table}{0}
\section{Exploratory Corpus of Prior Friction Designs}
\label{app:friction-corpus}

The friction designs included in our exploratory corpus are presented in \autoref{tab:friction-designs}. The corpus contains 73 unique designs from 24 papers published within the past five years. Four designs instantiate two mechanisms and are therefore listed under both categories, resulting in 77 design-mechanism entries.

\begingroup
\small
\renewcommand{\arraystretch}{1.2}
\setlength{\tabcolsep}{5pt}

\begin{longtable}[c]{
    >{\raggedright\arraybackslash}p{0.20\textwidth}
    >{\raggedright\arraybackslash}p{0.20\textwidth}
    >{\raggedright\arraybackslash}p{0.52\textwidth}
}

\caption{Prior friction designs examined in deriving the three friction mechanisms.}
\label{tab:friction-designs}\\

\toprule
\textbf{Mechanism}
& \textbf{Paper}
& \textbf{Design description}\\
\midrule
\endfirsthead

\multicolumn{3}{l}{
    \small\textit{Table \thetable\ continued from the previous page.}
}\\
\toprule
\textbf{Mechanism}
& \textbf{Paper}
& \textbf{Design description}\\
\midrule
\endhead

\midrule
\multicolumn{3}{r}{
    \small\textit{Continued on the next page.}
}\\
\endfoot

\bottomrule
\endlastfoot

\multirow{2}{*}{\underline{\textit{Eliciting user contribution}}} 
& \citet{haliburton2024longitudinal} & At each phone unlock, the \textit{Rabbit-Hole-Tracker} asks users to state
their intended purpose for using the phone. When the phone is locked, the app
asks whether they completed or exceeded that intention and whether they
regretted any part of the session.
\\
\addlinespace

& \citet{dalsgaard2025designing} & Proposed micro-prompt which asks users to reconsider whether an AI-generated suggestion reflects their intention or how they would express it themselves.
\\
\addlinespace

& \citet{dalsgaard2025designing} & Proposed attribution interface which asks users to label segments according to how human and AI contributions shaped them, such as generated, edited, or jointly constructed.
\\
\addlinespace

& \citet{dalsgaard2025designing} & Proposed reflective prompt which highlights repeated stylistic choices or culturally situated language and invites users to reconsider or explore another direction.
\\
\addlinespace

& \citet{buccinca2021trust} & AI-assisted decision interface which requires users to make an initial unaided decision before revealing the model suggestion and explanation, after which they may revise their decision.
\\

\addlinespace

& \citet{pennycook2021shifting} & Accuracy prompt which asks users to rate the accuracy of a single unrelated news headline before they make subsequent sharing decisions.
\\
\addlinespace

& \citet{joshi2026writing} & Prompt entry interface which requires users with shorter prompts to press and hold the submission button for longer, while writing more words reduces or removes the delay.
\\
\addlinespace

& \citet{joshi2026writing} & Prompt entry interface which requires users with shorter prompts to repeatedly move a slider to fill a submission gauge, while writing more words reduces the required movements.
\\
\addlinespace

& \citet{joshi2026writing} & Prompt entry interface which displays AI-generated questions after users pause, suggesting ways to expand or add detail to their prompt before submission.
\\
\addlinespace

& \citet{cai2024antagonisticai} & Conversational game which delivers personal insults and requires users to creatively agree with and respond to them as a form of guided self-reflection.
\\
\addlinespace

& \citet{cai2024antagonisticai} & Proposed AI intervention which detects potentially poor decisions and initiates an antagonistic dialogue asking users to explain and reconsider their intended action.
\\

\addlinespace

& \citet{lyngs2020just} & Facebook extension which asks users to state their purpose when opening the site and periodically displays that purpose in an expanding reminder until dismissed.
\\

\addlinespace

& \citet{10.1145/3800645.3813054} & Card-based toolkit which asks designers to juxtapose human values with potential harms and use reflective prompts to connect these tensions to concrete AI design choices.
\\
\addlinespace

& \citet{10.1145/3800645.3813054} & Spatial mapping canvas which requires designers to place two Value–Harm Cards along its axes and position AI concepts within the resulting quadrants to make trade-offs explicit and explore alternative designs.
\\
\addlinespace

& \citet{joshi2025writing} & Prompt submission interface where shorter prompts require a longer press-and-hold delay, encouraging users to write more before submitting. 
\\
\addlinespace

& \citet{kazemitabaar2025exploring} & The Guided-Write-Over interface presents AI-generated code as ghost text that users must type over line by line before they can use it, with explanations displayed as they type. 
\\
\addlinespace

& \citet{kazemitabaar2025exploring} & The Explain-before-Usage interface displays AI-generated code but requires users to answer questions explaining highlighted sections before they can use it.
\\
\addlinespace

& \citet{kazemitabaar2025exploring} & The Trace-and-Predict interface lets users step through AI-generated code execution but pauses at key points until they predict variable values. 
\\
\addlinespace

& \citet{ruiz2024design} & Social media feed which requires users to select a reaction or indicate that they are not interested before revealing the next post.
\\
\addlinespace

& \citet{10.1145/3479600} & Proposed prompt which, after detecting prolonged or unintended use, asks users to recall what they were doing before opening the application or what they intended to do afterward.
\\

\addlinespace

& \citet{i̇nan2025betterslowsorryintroducing} & Dialogue behavior which asks users questions about the context, previous statements, goals, or future steps before proceeding.
\\
\addlinespace

& \citet{zhang2025friction} & Reflective planning interface which requires users to manually cluster related feedback, diagnose the underlying writing problem, and formulate a revision strategy before revising the text, with optional AI hints available when needed.
\\
\addlinespace

& \citet{zhang2025friction} & Iterative revision interface which evaluates each user-written revision, explains why it is or is not an improvement, and prompts users to refine the sentence through additional attempts.
\\
\midrule

\multirow{2}{*}{\underline{\textit{Restricting access}}} 
& \citet{10.1145/3591156.3591183} & Time bar which delays action completion, creating a brief opportunity for reflection or undo.
\\
\addlinespace

& \citet{kazemitabaar2025exploring} & The Lead-and-Reveal interface which withholds AI-generated code and progressively reveals each line only after users reason about the next step in the solution. 
\\

\addlinespace

& \citet{collins2024modulatinglanguagemodelexperiences} & Selective-access interface which requires users to confirm through a second button before viewing the model prediction and reminds them of their estimated performance relative to the model.
\\
\addlinespace

& \citet{10.1145/3591156.3591183} & Pop-up which interrupts action completion, presents the consequences of the pending decision, and requires users to choose whether to revise or continue. 
\\
\addlinespace

& \citet{10.1145/3604241} & Interface which gradually obscures or shuts down the screen after detecting prolonged unintended use, giving users time to prepare to end the interaction. 
\\
\addlinespace

& \citet{ruiz2024design} & Social media feed which requires users to select a reaction or indicate that they are not interested before revealing the next post.
\\
\addlinespace

& \citet{10.1145/3604241} & Interface which interrupts unintended use and requires an explicit user action to continue. 
\\
\addlinespace

& \citet{10.1145/3591156.3591183} & Pop-up which interrupts action completion, presents the consequences of the pending decision, and requires users to choose whether to revise or continue. 
\\
\addlinespace

& \citet{10.1145/3491102.3517722} & Time-limit dialog which interrupts the interaction after 20 minutes and asks users whether to exit or continue using the application.
\\
\addlinespace

& \citet{10.1145/3479600} & Proposed intervention which limits the number of recommended posts or videos available within a given period to interrupt continued consumption.
\\
\addlinespace

& \citet{haliburton2024longitudinal} & Intervention which guides users through a brief breathing animation before allowing them to continue to the target application.
\\
\addlinespace

& \citet{haliburton2024longitudinal} & Intervention which presents a simplified breathing animation during a brief delay before application access.
\\
\addlinespace

& \citet{haliburton2024longitudinal} & Intervention which requires users to follow a moving dot with their finger before continuing to the target application.
\\
\addlinespace

& \citet{haliburton2024longitudinal} & Intervention which requires users to rotate their phone three times before continuing to the target application.
\\
\addlinespace

& \citet{haliburton2024longitudinal} & Intervention which displays the user’s face through the front-facing camera during a brief delay before application access.
\\
\addlinespace

& \citet{haliburton2024longitudinal} & Intervention which displays a blank screen during a brief delay before allowing access to the target application.
\\

\addlinespace

& \citet{jahn2023friction} & Proposed social media prompt which presents users with a short quiz about platform community standards before they can like or share a post.
\\
\addlinespace

& \citet{Cho26072024} & Online community which initially hides higher-risk features and progressively unlocks them only after users demonstrate supportive and helpful participation recognized by other community members.
\\
\addlinespace

& \citet{buccinca2021trust} & AI-assisted decision interface which hides the model suggestion and explanation by default and reveals them only when users explicitly request them.
\\
\addlinespace

& \citet{buccinca2021trust} & AI-assisted decision interface which requires users to make an initial unaided decision before revealing the model suggestion and explanation, after which they may revise their decision.
\\
\addlinespace

& \citet{buccinca2021trust} & AI-assisted decision interface which delays the model suggestion and explanation for 30 seconds, giving users time to form their own hypothesis before the output is revealed.
\\
\addlinespace

& \citet{dalsgaard2025designing} & Proposed interface which briefly delays the display or selection of AI-generated content, or greys out options until users have reviewed them.
\\
\midrule

\multirow{2}{*}{\underline{\textit{Reshaping system output}}}
& \citet{kazemitabaar2025exploring} & The Solve-Code-Puzzle interface transforms AI-generated code into scrambled draggable blocks that users must rearrange into a correct program before use. 
\\
\addlinespace

& \citet{kazemitabaar2025exploring} & The Verify-and-Review interface introduces errors into AI-generated code and requires users to identify and repair them before using the code. 
\\
\addlinespace

& \citet{kazemitabaar2025exploring} & The Interactive Pseudocode interface provides hierarchical pseudocode instead of a complete code solution and requires users to implement the corresponding code with guided feedback. 
\\

\addlinespace

& \citet{10.1145/3491102.3517722} & Feed filter which allows users to hide selected types of posts, such as replies, reposts, or original posts.
\\
\addlinespace

& \citet{10.1145/3491102.3517722} & Recommended-post blocker which removes algorithmically recommended content from the default feed.
\\
\addlinespace

& \citet{10.1145/3491102.3517722} & Custom lists which organize posts from selected accounts into separate user-curated feeds.
\\
\addlinespace

& \citet{lukoff2023switchtube} & YouTube client which allows users to switch between a recommendation-first Explore Mode and a search-first Focus Mode that hides homepage recommendations and removes related videos and autoplay.
\\
\addlinespace

& \citet{i̇nan2025betterslowsorryintroducing} & Dialogue behavior which explicitly states assumptions about the context, prior conversation, or the interlocutors’ reasoning so that they can be examined or corrected.
\\
\addlinespace

& \citet{i̇nan2025betterslowsorryintroducing} & Dialogue behavior which provides additional, unrequested details or explanations that may help clarify actions, choices, or beliefs.
\\
\addlinespace

& \citet{i̇nan2025betterslowsorryintroducing} & Dialogue behaviour which repeats or restates previously communicated information to emphasize or confirm it.
\\
\addlinespace

& \citet{subramonyam2024bridging} & Interface which presents multiple model outputs for the same prompt, allowing users to compare alternatives and relate differences in their prompts to differences in the generated results.
\\
\addlinespace

& \citet{subramonyam2024bridging} & Interface which augments model outputs with explanations, inline comments, summaries, or keywords to make them easier to interpret and evaluate.
\\

\addlinespace

& \citet{gosline2024nudge} & LLM output interface which uses color-coded highlighting to mark information as likely correct, potentially incorrect, or omitted, directing users to scrutinize and verify the generated text.
\\
\addlinespace

& \citet{cai2024antagonisticai} & AI chatbot which responds to every user idea with deliberately harsh criticism and devil’s-advocate feedback rather than supportive assistance.
\\
\addlinespace

& \citet{cai2024antagonisticai} & Conversational game which delivers personal insults and requires users to creatively agree with and respond to them as a form of guided self-reflection.
\\
\addlinespace

& \citet{cai2024antagonisticai} & Proposed smart-glasses assistant which delivers critical comments about users' clothing, language, and posture through bone-conduction audio.
\\
\addlinespace

& \citet{cai2024antagonisticai} & Proposed role-playing assistant which simulates hostile or negative reactions to help users rehearse difficult coming-out conversations.
\\
\addlinespace

& \citet{cai2024antagonisticai} & Proposed adaptive assistant which provides either supportive or antagonistic feedback based on the user's inferred emotional state.
\\
\addlinespace

& \citet{cai2024antagonisticai} & Proposed professional-feedback system which replaces softened workplace feedback with deliberately blunt and antagonistic critiques intended to motivate improvement.
\\
\addlinespace

& \citet{dalsgaard2025designing} & Proposed generative interface which presents divergent outputs side by side, preserves branching histories, and allows users to fork and revisit alternative creative directions.
\\

\midrule

\multirow{2}{*}{\underline{\textit{Other}}}
& \citet{kazemitabaar2025exploring} & The Solve-Code-Puzzle interface transforms AI-generated code into scrambled draggable blocks that users must rearrange into a correct program before use. 
\\
\addlinespace

& \citet{dalsgaard2025designing} & Proposed interface which visually distinguishes human- and AI-generated content and allows users to inspect the generative pathway that produced an output.
\\
\addlinespace

& \citet{dalsgaard2025designing} & Proposed interface which annotates AI outputs with information about their training-data provenance, dominant cultural patterns, or model limitations at the point of use.
\\
\addlinespace

& \citet{10.1145/3604241} & Timer which communicates how long unintended use has continued, either floating above the interface or blended into the displayed content. 
\\
\addlinespace

& \citet{10.1145/3604241} & Prompt which asks users to reflect on their emotional state, recall their initial intention, and indicate whether that intention has been fulfilled. 
\\
\addlinespace

& \citet{10.1145/3491102.3517722} & Reading progress indicator which marks the boundary between new and previously viewed posts and informs users when they have exhausted new content.
\\

\addlinespace

& \citet{10.1145/3479600} & Proposed content-status indicator which previews how much new content is available before users open or revisit an application.
\\
\addlinespace

& \citet{10.1145/3479600} & Proposed indicator which estimates how long users are likely to spend after entering a feature, making the prospective time cost visible before continued use.
\\
\addlinespace

& \citet{i̇nan2025betterslowsorryintroducing} & Dialogue behavior which introduces a verbal, embodied, or plan-recalibrating pause to signal reflection before the interaction continues.
\\
\addlinespace

& \citet{subramonyam2024bridging} & Node-based interface which visually traces prompts, outputs, and branching iterations, allowing users to review alternative paths and revise subsequent prompts.
\\
\addlinespace

& \citet{lyngs2020just} & Facebook extension which asks users to state their purpose when opening the site and periodically displays that purpose in an expanding reminder until dismissed.
\\

\addlinespace

& \citet{Cho26072024} & Online community which repeatedly reminds users during platform interactions not to disclose information that could reveal their identities.
\\

\end{longtable}
\endgroup

\section{Writing Prompts}
\label{apx:prompts}
\begin{enumerate}[label=\arabic*.]
\item Write about the advantages and disadvantages of using artificial intelligence in environmental protection.
\item Write about whether nuclear power should be used to meet future energy needs.
\item Write about the benefits and risks of genetic modification of crops.
\item Write about whether automation will create more jobs or eliminate them.
\item Write about the pros and cons of universal basic income.
\item Write about whether space exploration spending is justified when poverty still exists.
\item Write about the role of creativity versus academic knowledge in modern education.
\item Write about the effects of facial recognition technology in public spaces.
\end{enumerate}

\section{Semi-structured Interview Questions}
\label{apx:questions}
\begin{enumerate}[label=\arabic*.]
\item How would you describe your overall experience when using the different models to complete the writing tasks?

\item Were there any models that you found particularly challenging to use? 

\item Were there any models that you found particularly easy to use? 

\item In the study you were asked questions about ownership; how would you define ownership?

\item Which model, if any, made you feel the greatest sense of ownership over your writing?

\item If you had the opportunity to design your own writing model, what features or capabilities would you include?
\end{enumerate}

\section{Recall Scoring Pipeline}
\label{tab:semantic-scoring}
\begin{table*}[h]
\centering
\caption{Recall Scoring Examples.}

\small
\renewcommand{\arraystretch}{1.2} 

\begin{tabular}{p{5cm} p{7cm} >{\raggedleft\arraybackslash}p{1.2cm}}
\toprule
\textbf{Recalled Text} & \textbf{Written Text} & \textbf{Score} \\
\midrule

\multirow[t]{5}{5cm}{\raggedright
Nuclear power should be used to meet future energy needs. Natural resources are not very reliable as they are not always available (eg. wind power, solar power); however, nuclear power uses (some element I forget) which is abundantly available. It is really important to be cautious and careful with nuclear waste disposal.
}
& I think nuclear power should be used to meet our future energy needs. & 1 \\
& Using natural resources can be challenging as they are not always readily available (eg. wind power or solar). & 1 \\
& However, nuclear power is some relies on uranium which is quite abundant in quantity. & 0.5 \\
& This makes it a promising resource to use for future energy needs! & 0 \\
& It is important to be mindful of careful and cautious disposal though. & 1 \\

\bottomrule
\end{tabular}

\end{table*}

\section{Per-intervention Scores}
\label{app:workload}

\begin{table}[H]
\centering
\caption{Comparison of overall workload scores for each intervention versus the ChatGPT control.} 
\small

\begin{tabular}{lccc}
  \toprule
  \textbf{Intervention} & $\boldsymbol{\Delta M}$ & \textbf{95\% CI} & \textbf{\textit{p}} \\
  \midrule
  \textsc{Think Multiplier}     & 5.73 & [4.18, 7.40] & $< .001$ \\
  \textsc{Timer Lockout}        & 5.65 & [3.97, 7.68] & $< .001$ \\
  \textsc{Elenchus}             & 5.59 & [3.99, 7.36] & $< .001$ \\
  \textsc{Less is More}         & 4.56 & [3.00, 6.71] & $< .001$ \\
  \textsc{Hold Reveal}          & 4.55 & [3.06, 6.07] & $< .001$ \\
  \textsc{Write by Yourself}    & 3.44 & [2.11, 4.67] & $< .001$ \\
  \textsc{User Echo}            & 2.13 & [0.67, 3.61] & $.007$ \\
  \bottomrule
\end{tabular}

\end{table}

\begin{table*}[h]
\centering
\caption{NASA-TLX dimension-level effects across interventions versus ChatGPT.} 
\label{tab:workload_subscales}
\small

\sisetup{
  detect-all,
  input-symbols = {()},
  table-number-alignment = center,
  round-mode = places,
  round-precision = 2
}

\begin{tabular}{ll S[table-format=2.2] l l}
\toprule
\textbf{Intervention} & \textbf{Dimension} & $\boldsymbol{\Delta M}$ & \textbf{95\% CI} & {\textbf{\textit{p}}} \\
\midrule

\multirow{5}{*}{\textsc{Think Multiplier}}
  & Mental Demand   & 7.33 & {[4.92, 9.75]} & <.001 \\
  & Physical Demand & 3.29 & {[1.75, 5.04]} & .002 \\
  & Effort          & 8.75 & {[6.54, 10.83]} & <.001 \\
  & Performance     & 1.67 & {[0.04, 3.42]} & 0.84 \\
  & Frustration     & 7.58 & {[5.21, 9.96]} & <.001 \\
\addlinespace

\multirow{5}{*}{\textsc{Timer Lockout}}
  & Mental Demand   & 6.625 & {[4.38, 8.92]} & <.001 \\
  & Physical Demand & 5.63 & {[3.38, 8.04]} & .001 \\
  & Effort          & 7.54 & {[5.38, 9.75]} & <.001 \\
  & Performance     & 0.92 & {[-0.71, 2.54]} & 0.84 \\
  & Frustration     & 7.54 & {[5.04, 10.04]} & <.001 \\
\addlinespace

\multirow{5}{*}{\textsc{Elenchus}}
  & Mental Demand   & 8.00 & {[5.79, 10.17]} &  <.001\\
  & Physical Demand & 0.00 & {[7.25, 11.125]} & .011 \\
  & Effort          & 9.21 & {[7.25, 11.125]} & <.001 \\
  & Performance     & 1.33 & {[0.04, 2.63]} & 0.61 \\
  & Frustration     & 6.54 & {[4.33, 8.92]} & <.001 \\
\addlinespace

\multirow{5}{*}{\textsc{Less is More}}
  & Mental Demand   & 5.54 & {[3.125, 7.96]} & .002 \\
  & Physical Demand & 3.58 & {[1.54, 5.88]} & .011 \\
  & Effort          & 6.88 & {[4.92, 8.83]} & <.001 \\
  & Performance     & 0.96 & {[-0.75, 3]} & 1 \\
  & Frustration     & 5.83 & {[3.29, 8.54]} & .003 \\
\addlinespace

\multirow{5}{*}{\textsc{Hold Reveal}}
  & Mental Demand   & 4.34 & {[2.48, 6.43]} & .002 \\
  & Physical Demand & 5.69 & {[3.74, 7.78]} & .001 \\
  & Effort          & 5.91 & {[3.70, 8.09]} & <.001 \\
  & Performance     & 0.96 & {[-0.26, 2.35]} & 1 \\
  & Frustration     & 5.74 & {[3.57, 7.91]} & .002 \\
\addlinespace

\multirow{5}{*}{\textsc{Write by Yourself}}
  & Mental Demand   & 6.71 & {[4.79, 8.63]} & <.001 \\
  & Physical Demand & 2.25 & {[0.79, 4.04]} & .01 \\
  & Effort          & 6.13 & {[3.92, 8.21]} & <.001 \\
  & Performance     & 0.42 & {[-1.46, 2.08]} & 1 \\
  & Frustration     & 1.71 & {[-0.42, 3.79]} & .05 \\
\addlinespace

\multirow{5}{*}{\textsc{User Echo}}
  & Mental Demand   & 3.42 & {[1.75, 5.13]} & .002 \\
  & Physical Demand & 1.13 & {[0.04, 2.42]} & .06 \\
  & Effort          & 3.96 & {[2.21, 5.67]} & <.001 \\
  & Performance     & -0.17 & {[-1.83, 1.42]} & 1 \\
  & Frustration     & 2.29 & {[0.08, 4.42]} & .05 \\
\addlinespace

\bottomrule
\end{tabular}

\end{table*}

\begin{figure*}[h]
\centering
\includegraphics[width=\twocolwidth]{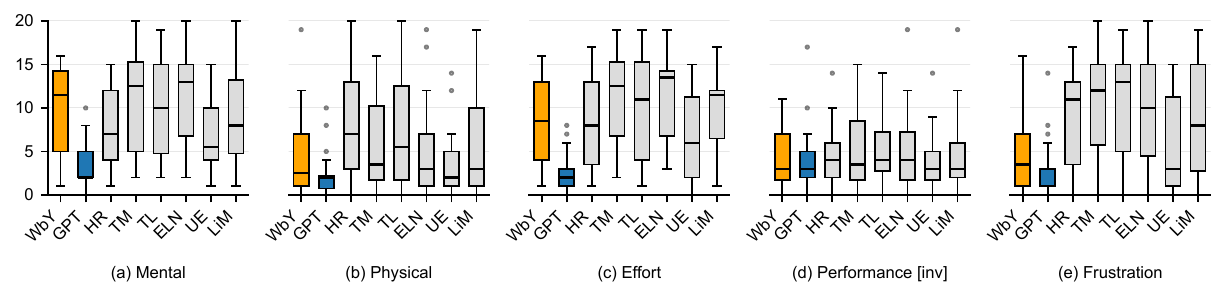}
\caption{NASA\_TLX scores by interface. Abbreviations for conditions: \textsc{Write by Yourself} (WbY), \textsc{ChatGPT} (GPT), \textsc{Hold Reveal} (HR), \textsc{Think Multiplier} (TM), \textsc{Timer Lockout} (TL), \textsc{Elenchus} (ELN), \textsc{User Echo} (UE), and \textsc{Less is More} (LiM). We highlight \textsc{Write by Yourself} in yellow and \textsc{ChatGPT} conditions in blue; all friction conditions are in grey.}
\label{fig:subworkload} 
\end{figure*}

\begin{table}[H]
\centering
\caption{Comparison of ownership scores for each intervention versus the ChatGPT control.} 
\small

\begin{tabular}{lccc}
  \toprule
  \textbf{Intervention} & $\boldsymbol{\Delta M}$ & \textbf{95\% CI} & \textbf{\textit{p}} \\
  \midrule
  Write by Yourself & 2.71 & [2.10, 3.21] & $< .001$ \\
Elenchus & 2.25 & [1.59, 2.79] & $< .001$ \\
Think Multiplier & 1.92 & [1.27, 2.50] & $< .001$ \\
Less Is More & 1.38 & [0.83, 1.85] & $.001$ \\
Timer Lockout & 1.21 & [0.58, 1.81] & $.005$ \\
User Echo & 1.04 & [0.44, 1.72] & $.01$ \\
Hold Reveal & 1.00 & [0.27, 1.64] & $.01$ \\
  \bottomrule
\end{tabular}

\end{table}

\begin{table}[H]
\centering
\caption{Comparison of task duration for each intervention versus the ChatGPT control.} 
\small

\begin{tabular}{lccc}
  \toprule
  \textbf{Intervention} & $\boldsymbol{\Delta M}$ & \textbf{95\% CI} & \textbf{\textit{p}} \\
  \midrule
  Think Multiplier & 3.41 & [2.32, 4.90] & $< .001$ \\
Elenchus & 2.96 & [2.03, 3.91] & $< .001$ \\
User Echo & 2.26 & [1.26, 3.59] & $< .001$ \\
Less Is More & 2.16 & [1.14, 3.26] & $< .001$ \\
Timer Lockout & 2.03 & [1.19, 2.76] & $< .001$ \\
Hold Reveal & 2.00 & [1.12, 3.14] & $< .001$ \\
Write by Yourself & 1.33 & [0.76, 2.06] & $< .001$ \\
  \bottomrule
\end{tabular}

\end{table}

\begin{table}[H]
\centering
\caption{Comparison of recall performance for each intervention versus the ChatGPT control.} 
\small

\begin{tabular}{lccc}
  \toprule
  \textbf{Intervention} & $\boldsymbol{\Delta M}$ & \textbf{95\% CI} & \textbf{\textit{p}} \\
  \midrule
  Write by Yourself & 0.50 & [0.34, 0.66] & $< .001$ \\
Elenchus & 0.40 & [0.24, 0.56] & $.002$ \\
Timer Lockout & 0.29 & [0.11, 0.48] & $.02$ \\
Think Multiplier & 0.23 & [0.09, 0.41] & $.03$ \\
Hold Reveal & 0.23 & [0.08, 0.38] & $.02$ \\
Less Is More & 0.21 & [0.05, 0.37] & $.03$ \\
User Echo & 0.07 & [-0.07, 0.24] & $.07$ \\
  \bottomrule
\end{tabular}

\end{table}

\begin{figure*}[h]
\centering
\includegraphics[width=0.5\twocolwidth]{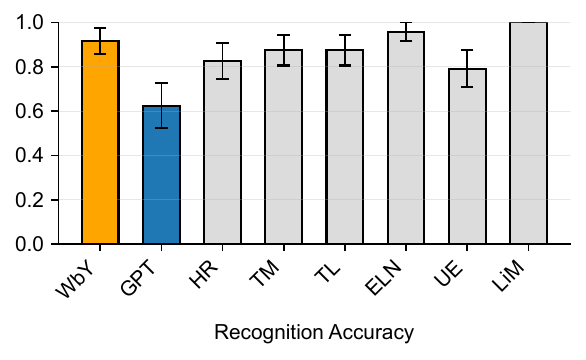}
\caption{Participants’ recognition accuracy across writing interfaces. Abbreviations for conditions: \textsc{Write by Yourself} (WbY), \textsc{ChatGPT} (GPT), \textsc{Hold Reveal} (HR), \textsc{Think Multiplier} (TM), \textsc{Timer Lockout} (TL), \textsc{Elenchus} (ELN), \textsc{User Echo} (UE), and \textsc{Less is More} (LiM). Recognition accuracy is presented as a bar chart representing the mean score, due to the binary nature of the metric (0 = incorrect, 1 = correct). Error bars indicated standard error.}
\label{fig:recognition} 
\end{figure*}

\clearpage
\section{\textsc{Usage Mirror}}
\label{tab:usage_mirror}
\begin{figure*}[h]
	\centering
	\includegraphics[width=\twocolwidth]{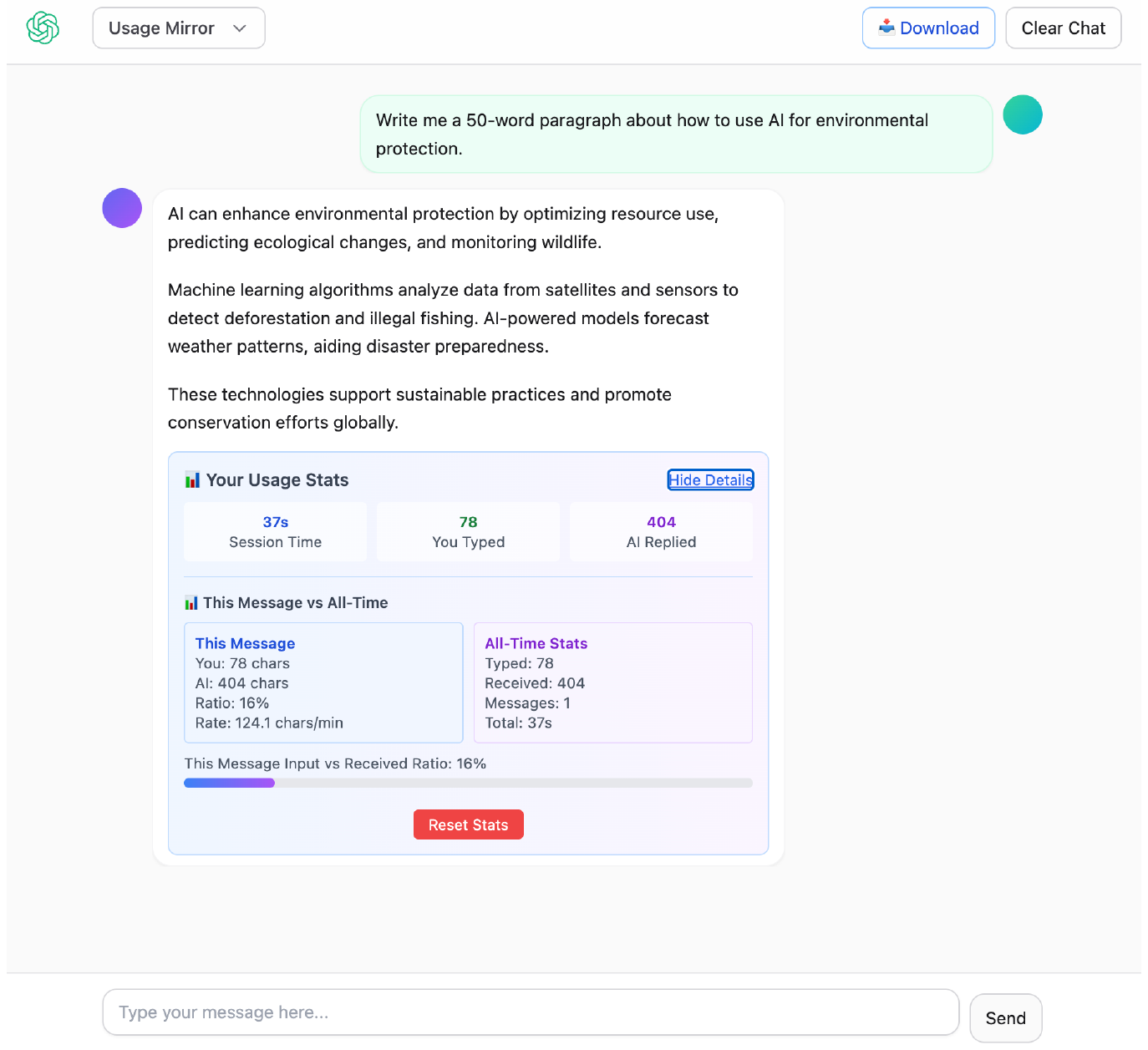}
	\caption{The \textsc{Usage Mirror} interface. This design introduces friction by visualizing real-time interaction metrics, including session duration and the character count ratio between user input and LLM output (bottom panel). It is intended to promote self-monitoring and voluntary self-correction without obstructing the workflow.}
\end{figure*}

\end{document}